\PassOptionsToPackage{dvipsnames}{xcolor}
\documentclass[10pt, a4paper, twocolumn]{article}

\usepackage{graphicx}
\usepackage{xcolor}
\usepackage[utf8]{inputenc}
\usepackage[english]{babel}
\usepackage[autostyle=true]{csquotes}
\usepackage{url}
\usepackage{tabularx}
\usepackage{multirow}
\usepackage{multicol}
\usepackage{booktabs}
\usepackage{listings}
\usepackage{enumitem}
\usepackage{amsfonts}
\usepackage{balance}
\usepackage{adjustbox}
\usepackage{hyperref}
\usepackage{amsmath}
\usepackage[labelfont={bf}]{caption}
\usepackage{graphicx}
\usepackage{textcomp}
\usepackage{xcolor}
\usepackage{listingsutf8}
\usepackage[all]{nowidow}
\usepackage[capitalise,noabbrev]{cleveref}
\usepackage{subcaption}
\usepackage{authblk}
\usepackage{pdfpages}

\usepackage{todonotes}

\date{}
\begin{document}
	\null
	\includepdf[lastpage=1,pages=-, scale=1.0]{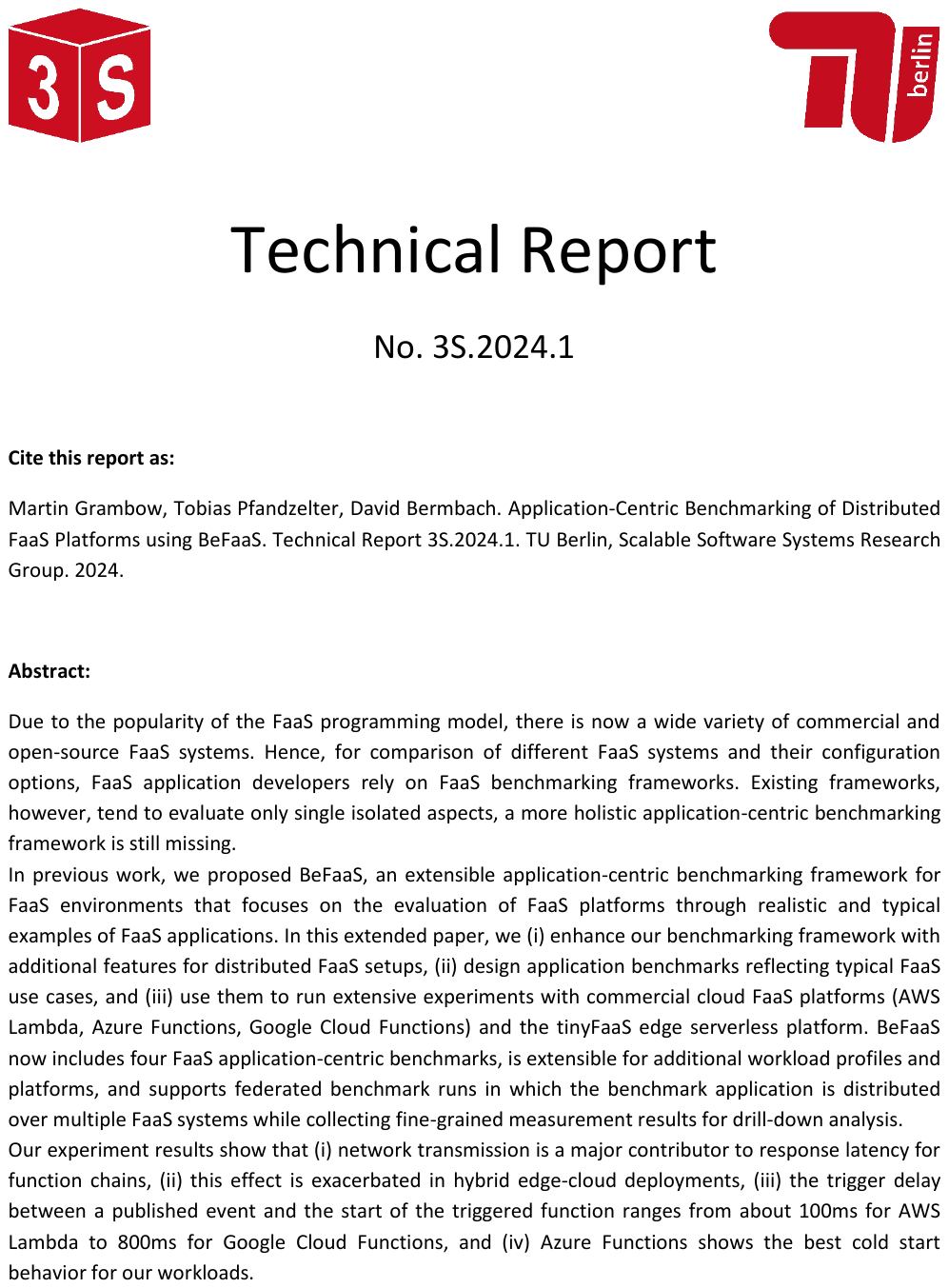}

		\title{Application-Centric Benchmarking of Distributed FaaS Platforms using BeFaaS\thanks{This work extends~\cite{grambow_befaas_2021}.}}

		\author{Martin~Grambow$^1$,
				Tobias Pfandzelter$^1$,
				and~David~Bermbach$^1$}
						
		\affil{$^1$Mobile Cloud Computing Research Group, TU Berlin \\Berlin, Germany. \\E-mail: \{mg,tp,db\}@mcc.tu-berlin.de}

		\maketitle

    \begin{abstract}
        Due to the popularity of the FaaS programming model, there is now a wide variety of commercial and open-source FaaS systems.
		Hence, for comparison of different FaaS systems and their configuration options, FaaS application developers rely on FaaS benchmarking frameworks.
		Existing frameworks, however, tend to evaluate only single isolated aspects, a more holistic application-centric benchmarking framework is still missing.

        In previous work, we proposed BeFaaS, an extensible application-centric benchmarking framework for FaaS environments that focuses on the evaluation of FaaS platforms through realistic and typical examples of FaaS applications.
        In this extended paper, we (i) enhance our benchmarking framework with additional features for distributed FaaS setups, (ii) design application benchmarks reflecting typical FaaS use cases, and (iii) use them to run extensive experiments with commercial cloud FaaS platforms (AWS Lambda, Azure Functions, Google Cloud Functions) and the tinyFaaS edge serverless platform.
        BeFaaS now includes four FaaS application-centric benchmarks, is extensible for additional workload profiles and platforms, and supports federated benchmark runs in which the benchmark application is distributed over multiple FaaS systems while collecting fine-grained measurement results for drill-down analysis.

        Our experiment results show that (i) network transmission is a major contributor to response latency for function chains, (ii) this effect is exacerbated in hybrid edge-cloud deployments, (iii) the trigger delay between a published event and the start of the triggered function ranges from about 100ms for AWS Lambda to 800ms for Google Cloud Functions, and (iv) Azure Functions shows the best cold start behavior for our workloads.
    \end{abstract}
		
		\vspace{1em}


\section{Introduction}

All major cloud providers offer Function-as-a-Service (FaaS) solutions where users only have to take care of their source code (functions) while the underlying infrastructure and environment are abstracted away by the provider.
FaaS applications are composed of individual functions deployed on a FaaS platform that handles, e.g., the execution and automatic scaling.
Developers do not have direct control of the infrastructure and can only define high-level parameters, such as the region in which the function should run~\cite{paper_bermbach_cloud_engineering}.
Due to this, FaaS platforms are easy to use but comparing cloud platform performance~\cite{paper_leitner_cloud_variability,paper_bermbach_expect_the_unexpected} is challenging, as the cloud variability is further compounded by an additional, unknown infrastructure component.

Existing work on benchmarking of FaaS platforms usually focuses on the execution of small, isolated microbenchmarks that deploy and call a single function, e.g., a matrix multiplication~\cite{back_using_2018} or a random number generator~\cite{malawski_benchmarking_2017}.
While \emph{microbenchmarks} are useful for studying and comparing specific characteristics, they can give only limited insights into the platform behavior that will impact real applications~\cite{book_cloud_service_benchmarking}.
An \emph{application-centric benchmark}, in contrast, mimics the behavior of a realistic application in order to more realistically measure platform performance.
This allows developers to better compare different service options, a strategy also taken by the TPC benchmarks.\footnote{\url{https://www.tpc.org}}

In previous work, we proposed \emph{BeFaaS}~\cite{grambow_befaas_2021}, an extensible framework for executing application-centric benchmarks against FaaS platforms that included a realistic e-commerce benchmark as an example, to address this gap.
At the time, BeFaaS was the only FaaS benchmarking framework with out-of-the-box support for federated cloud~\cite{paper_kurze_cloud_federation} and edge-to-cloud deployments~\cite{paper_bermbach2021_auctionwhisk,baresi2019towards}, which allowed us to evaluate complex application configurations distributed over platforms running on a mixture of cloud, edge, and fog nodes.
Beyond this, BeFaaS focuses on ease-of-use and collects fine-grained measurements which can be used for a detailed post-experiment drill-down analysis, e.g., to identify cold starts or trace request chains in detail.

In this extended version of our original paper~\cite{grambow_befaas_2021}, we enhance the BeFaaS framework with asynchronous cross-provider event pipelines, design and implement three further FaaS benchmark applications, add support for benchmarking edge-based FaaS platforms, and use BeFaaS to evaluate several FaaS providers in realistic and typical FaaS application setups.
Specifically, we (i) compare FaaS offerings using a typical microservice-based application, (ii) evaluate hybrid edge-cloud FaaS setups, (iii) analyze the event pipeline interaction within and between providers, and (iv) study cold start behavior of FaaS platforms.

In total, we make the following contributions:
\begin{itemize}
    \item We derive requirements for an application-centric FaaS benchmarking framework (\cref{sec:requirements}).
    \item We propose BeFaaS, an extensible framework for the execution of application-centric FaaS benchmarks and describe four example benchmark applications (\cref{sec:design}).
    \item We present our proof-of-concept prototype which is available as open source (\cref{sec:impl}).
    \item We run a number of experiments and use the results to compare FaaS offerings in several setups (\cref{sec:evaluation}).
\end{itemize}

Our extended study evaluates four different typical FaaS use cases on Amazon Web Services (AWS), Google Cloud Platform (GCP), Microsoft Azure (Azure), and tinyFaaS~\cite{pfandzelter_tinyfaas_2020}.
Our experiments result in the following findings:
\begin{enumerate}
    \item For simple functions which work as glue code between frontend and storage layer, network transmission is a major contributor to overall response latency while the pure computing time of functions is almost negligible.
    \item Even with a cloud database backend, an edge-only function deployment can outperform a mixed edge-cloud deployment in response latency as a result of transmission latency between functions.
    \item While publishing events to event pipelines can be done within 100ms for all studied providers, the delay between a published event and the start of the triggered function ranges between about 100ms for AWS and 800ms for GCP.
    \item Despite function execution duration being the highest on Azure Functions, our experiments find that the platform outperforms AWS Lambda and Google Cloud Functions in cold start behavior.
\end{enumerate}

BeFaaS can help FaaS application developers compare cloud offerings and find the best provider for their specific use case by either deriving findings from our benchmark applications, adjusting the workload profiles to match their scenario, or using the BeFaaS library in their specific FaaS application for most accurate findings.
Furthermore, FaaS platform developers could use BeFaaS as part of their CI/CD pipelines~\cite{waller2015including,grambow_continuous_benchmarking} to detect performance regressions prior to live testing.

\section{Related Work}
\label{sec:rw}

Existing research on benchmarking of FaaS environments has so far mostly focused on microbenchmarks.
Application-centric benchmarks that consider the overall performance of multiple functions, the interaction with external services, and the effects of different application load profiles are mostly still missing.

Microbenchmarks call single functions repeatedly and evaluate the resulting metrics.
These functions are often designed for a specific purpose, e.g., to stress the CPU of the test system or to evaluate the test system with a disk-intensive workload.
Most performance evaluation studies are based on microbenchmarks deployed on FaaS platforms of different vendors~\cite{wang_peeking_2018, malawski_benchmarking_2017, figiela_performance_2018, manner_cold_2018, shahrad_architectural_2019, lee_evaluation_2018, back_using_2018, yu_characterizing_2020, martins_benchmarking_2020}.
Besides scaling of functions, cold start latency, containerization overheads, and instance lifetimes, the studies also evaluate metrics such as CPU utilization, network throughput, and costs.
Almost all experiments, however, focus on a single isolated aspects and do not create a holistic comparability of platforms performance for FaaS application developers.

Several studies also consider more complex applications and focus on specific FaaS related features, e.g., by deploying image processing pipelines~\cite{kim_functionbench_2019}, analyzing chained functions, or deploying real world applications on serverless platforms~\cite{yu_characterizing_2020}.
While the authors of these studies also use application-centric workloads for experiments, their goal was not to propose a comprehensive framework for the execution of application-centric FaaS benchmarks.
Since the publication of the initial version of BeFaaS~\cite{grambow_befaas_2021}, recent research work has focused on benchmarking function triggers~\cite{scheuner2022triggerbench}, studying tail latency~\cite{ustiugov_analyzing_2021}, implications of used programming languages~\cite{cordingly_implications_2020}, hardware influence factors~\cite{cordingly2020predicting}, concurrent function executions~\cite{barcelonapons_benchmarking_2021}, take a closer look at the cost perspective~\cite{pfandzelter_streaming_2022}, performance fluctuations over time~\cite{schirmer_night_2023}, and general FaaS characteristics~\cite{eismann_state_2021}.

Further, there are several studies and frameworks that share some of the features and goals of BeFaaS:
\emph{PanOpticon}~\cite{somu_panopticon_2020} uses a deployment, workload, and metrics module to evaluate chained functions and a simple chat server on two different FaaS vendors.
Although PanOpticon has similar goals as BeFaaS, it neither supports detailed drill-down analysis nor federated multi-provider setups.
Van Eyk et al.~\cite{van_eyk_beyond_2020} develop a high-level architecture and state requirements for serverless benchmarking.
Unfortunately their project still appears to be in a vision state.
\emph{FaaSdom}~\cite{maissen_faasdom_2020} shares our motivation for a full application deployment.
It supports multiple platforms, several languages (e.g., Node.js, Python, Go), and an automatic deployment of performance tests via a web frontend.
All applications, however, use HTTP triggers and there are no function chains but single function applications which focus on different aspects such as CPU, latency, or IO performance similar to microbenchmarks.
\emph{SeBS}~\cite{copik_sebs_2021} is a FaaS benchmarking framework that highlights the cost efficiency of executions and, similar to FaaSdom, also only considers single-function applications.

Scheuner et al.~\cite{scheuner_lets_2022, scheuner_crossfit_2022} propose fine-grained tracing for serverless applications.
In contrast to BeFaaS, which uses the default logging mechanism of each individual provider to collect traces, they use the platforms' tracing frameworks.
This might lower the tracing overhead compared to writing traces to log files, yet it increases the difficulty of studying cross-provider traces, which BeFaaS supports out-of-the-box.
Besides a library which supports multi-cloud setups~\cite{zhao_supporting_2022}, to the best of our knowledge, BeFaaS is still the only FaaS benchmarking framework for evaluating federated cross-provider setups which can also be used to trace request in the edge to cloud continuum.

Beyond FaaS, there are a number of application-centric benchmarking frameworks in other domains, e.g., for database and storage systems~\cite{bermbach_benchfoundry_2017,difallah2013oltp} or for virtual machines~\cite{borhani2014wpress}. These can, however, not easily be adapted to FaaS platforms.

\section{Requirements}
\label{sec:requirements}

While microbenchmarks are highly useful for studying individual features of a system-under-test (SUT), application-centric benchmarks support end-to-end comparison of different platforms and configurations.
Aside from standard benchmarking requirements such as portability or fairness~\cite{huppler_art_2009, paper_bermbach_benchmarking_middleware,bermbach_benchfoundry_2017, folkerts_benchmarking_2013,book_cloud_service_benchmarking}, an application-centric FaaS benchmarking framework needs to fulfill a number of specific requirements which we describe in this section.

\paragraph*{R1 -- Realistic Benchmark Application:}
The performance of a FaaS platform depends on the application that is deployed on it.
For instance, an application that frequently causes cold starts through a growing request rate will be better off on AWS Lambda while an application that frequently causes cold starts through short temporary load spikes will be better off on Apache OpenWhisk due to their different request queuing mechanisms~\cite{paper_bermbach_faas_coldstarts}.
This means that the benchmark application should be as close as possible to the real application for which the analysis is made~\cite{book_cloud_service_benchmarking}, in line with the findings of Shahrad et al.~\cite{shahrad_serverless_2020}.
A key requirement is, hence, that \emph{a FaaS benchmark should mimic real applications as closely as possible}.

\paragraph*{R2 -- Extensibility for New Workloads:}
FaaS platforms are highly flexible and can be used for a wide variety of applications, so the world of FaaS applications is evolving rapidly.
As such, any set of ``typical'' FaaS applications -- and thus the workload profile for a FaaS platform -- can only be considered a snapshot in time.
Likewise, the load profiles of existing FaaS applications, i.e., the amount and type of requests that the application handles, are likely to evolve over time.
Therefore, we argue that \emph{a FaaS benchmarking framework should be easily extensible in terms of adding new benchmark applications and updating load profiles for existing benchmarks}.

\paragraph*{R3 -- Support for Modern Deployments:}
FaaS is often used as the ``glue'' between cloud services, web APIs, and legacy systems~\cite{paper_bermbach_cloud_engineering}.
Thus, a benchmarking framework must also consider these links and support external services.
Furthermore, applications today are often distributed over cloud, edge, and fog resources, possibly even to the LEO edge~\cite{paper_bermbach_fog_vision,paper_zhang_gdp, paper_pfandzelter_LEO_serverless}.
Here, for example, edge servers can keep sensitive functions on premises while non-critical functions are hosted in a public cloud; similar setups exist for edge and fog computing use cases~\cite{paper_pallas_fog4privacy,grambow_public_2018, aslanpour_serverless_nodate, paper_bermbach2021_auctionwhisk}.
As such, assuming a single-cloud deployment is unrealistic for benchmarks aiming to be as similar as possible to realistic applications.
\emph{A benchmarking framework needs to support external services and federated setups in which application functions are deployed on one or more FaaS platforms distributed across cloud, edge, and fog}.

\paragraph*{R4 -- Extensibility for New Platforms:}
Today, all major cloud service providers offer FaaS platforms and there is a growing range of open-source FaaS systems, e.g., systems that specifically target the edge~\cite{paper_george_nanolambda,pfandzelter_tinyfaas_2020}.
As interfaces are constantly evolving and new platforms are being introduced, a cross-platform benchmarking framework \emph{needs to be extensible to support future FaaS platforms}.

\paragraph*{R5 -- Support for Drill-down Analysis:}
An application-centric FaaS benchmark can help to evaluate the suitability of different sets and configurations of FaaS platforms for a specific application.
What it can usually not provide are explanations for its finding, e.g., the different cold start management behavior of AWS Lambda and Apache OpenWhisk mentioned above~\cite{paper_bermbach_faas_coldstarts}.
To facilitate root cause analysis and help evaluators explain the patterns they see in the benchmark results, we argue that \emph{an application-centric FaaS benchmarking framework should support drill-down analysis by logging fine-grained measurement results including typical metrics of microbenchmarks}.

\paragraph*{R6 -- Minimum Required Configuration Overhead:}
An application-centric FaaS benchmarking framework should be easy to use and provide reproducible results.
This includes configuration, deployment, execution, as well as collection and analysis of results, e.g., using infrastructure automation.
Hence, \emph{a FaaS benchmarking framework should be designed to require as little manual effort as possible}.
\section{Design}
\label{sec:design}
In this section, we give an overview of the BeFaaS design, starting with an overview of the BeFaaS architecture and components (\cref{subsec:architecture}) before describing the key features of BeFaaS (\crefrange{subsec:design1}{subsec:design-last}).

\begin{figure}
    \centering
    \includegraphics[width=0.95\columnwidth]{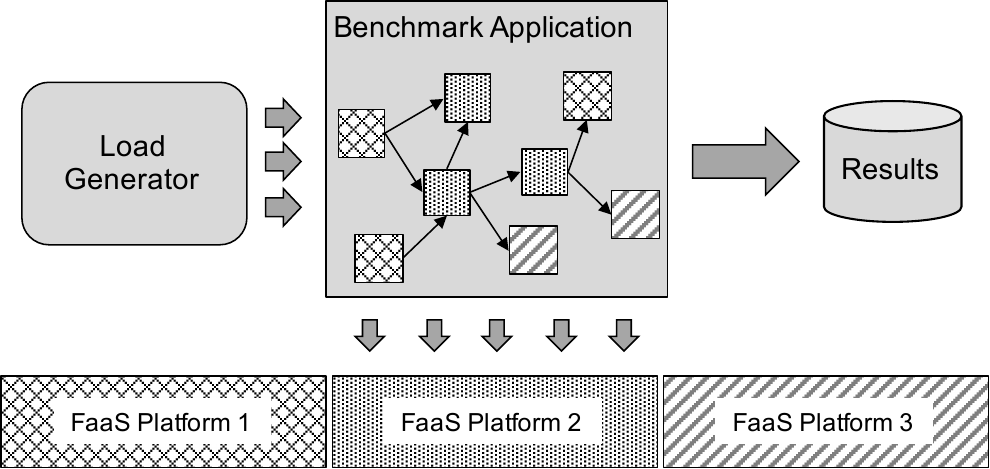}
    \caption{High-level overview of the BeFaaS architecture.}
    \label{fig:arch}
\end{figure}

\subsection{Architecture and Components\label{subsec:architecture}}
In BeFaaS, executing functions of a benchmark application is the \emph{workload} that actually benchmarks the FaaS platform, i.e., executing a function creates \emph{stress} on the SUT.
Since functions do not ``self-start'' executing, we need an additional load generator that invokes the FaaS functions of our benchmark application
We show a high-level architecture overview in \cref{fig:arch}.

\begin{figure}
    \centering
    \includegraphics[width=0.95\columnwidth]{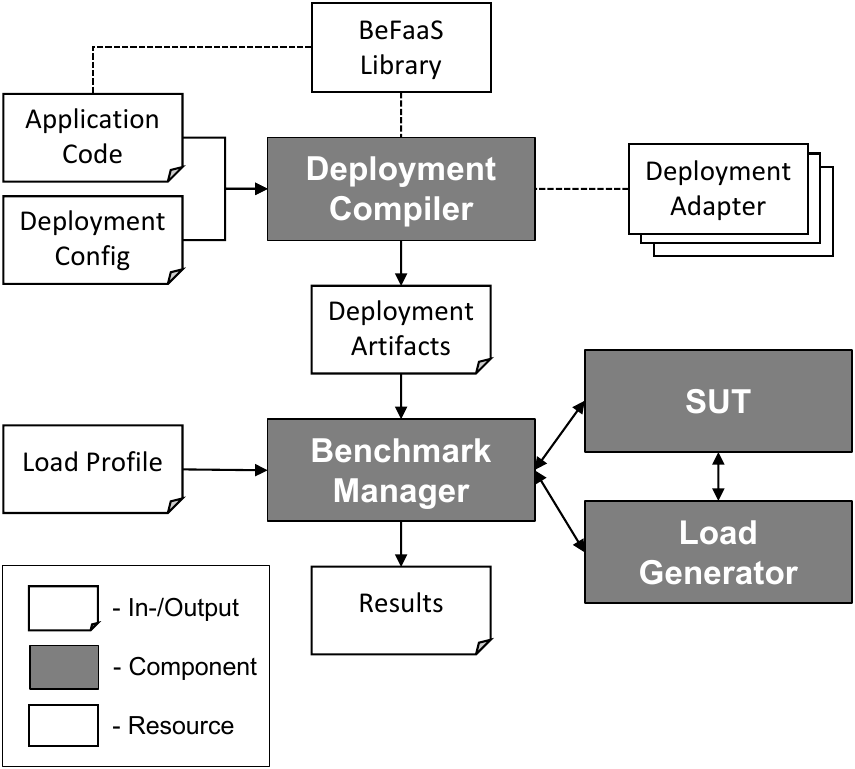}
    \caption{The Deployment Compiler transforms application code into individual deployment artifacts based on a deployment configuration. These are then deployed and invoked by the Load Generator to retrieve measurement results. Finally, the Benchmark Manager aggregates and reports fine-grained results.}
    \label{fig:overview}
\end{figure}

For a benchmark run, BeFaaS requires three inputs: (i)~the source code of the FaaS functions forming the benchmark application, (ii)~a load profile for the load generator, and (iii)~a deployment configuration that describes the environment configuration for each function and FaaS platform (the SUTs).
We show the components of BeFaaS and their interaction in \cref{fig:overview}.

Application code and deployment configuration are initially converted into deployment artifacts by the \emph{Deployment Compiler}.
The Deployment Compiler instruments and wraps each function's code with BeFaaS library calls and injects vendor-specific instructions defined in deployment adapters to enable request tracing and fine-grained metrics.
The resulting deployment artifacts are passed to the \emph{Benchmark Manager}.

The Benchmark Manager orchestrates the experiment:
First, it configures the \emph{SUT} by deploying each function based on the information in the respective artifact.
If there are external services, e.g., a database service, these can either be deployed by the Benchmark Manager as well or linked to the SUT using environment variables.
In the second step, the Benchmark Manager initializes the \emph{Load Generator} with the workload information described in a load profile.
Then, the benchmark run is triggered and the Load Generator invokes the functions of the benchmark application, which log every request in detail, including timestamps, origin function, and called functions (if applicable).
Once the benchmark run is completed the Benchmark Manager collects function logs, aggregates them, and destroys all provisioned resources.

\subsection{Realistic Benchmarks\label{subsec:design1}}
To provide a relevant and realistic application-centric benchmark (\textbf{R1}), BeFaaS already comes with four built-in benchmarks which mimic and represent typical use cases for FaaS applications.
These include a microservice-based web application to study request-response patterns, an IoT application scenario to evaluate hybrid edge-cloud setups, a smart factory application to measure event trigger performance, and a microservice application to study cold start behavior and elasticity capabilities (details are further explained in \cref{sec:impl}).
Our application benchmarks are in line with the empirical findings of Shahrad et al.~\cite{shahrad_serverless_2020} regarding typical FaaS applications:
All are composed of several functions that interact with each other to form function chains, use synchronous HTTP or asynchronous event triggers, and use external services such as a database system for persistence.
The benchmark applications come with a default load profile that covers all relevant aspects as well as several further load profiles to emphasize selected stress situations, e.g., to provoke more cold starts.
In combination, the benchmarks each represent complete FaaS applications: load balancing at the provider endpoint(s), interconnected calls of several functions, calls to external services such as database systems, and multiple load profiles which, e.g., provoke cold starts of functions.

The modular design of BeFaaS, however, also allows us to easily add further benchmark applications and load profiles or to adapt existing ones to the concrete needs of the developer (\textbf{R2}).
For adding a new benchmark, the respective application only needs to use the BeFaaS library (described in \cref{sec:impl}) for function calls and to have unique function names.

\begin{figure}
    \centering
    \includegraphics[width=0.7\columnwidth]{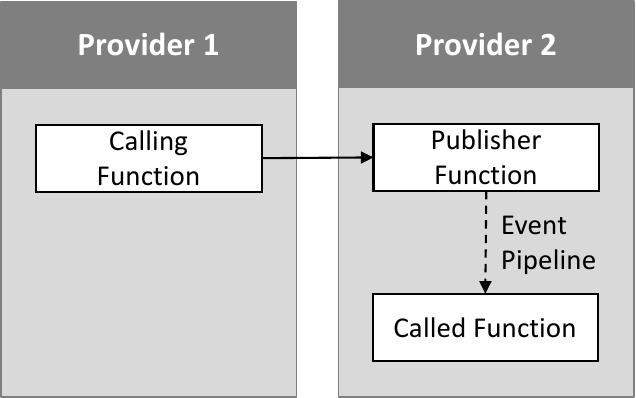}
    \caption{A publisher function forwards incoming events to the respective event pipeline to trigger the called function.}
    \label{fig:publisher}
\end{figure}

\subsection{Benchmark Portability and Federated FaaS Deployments}
To support portability of benchmarks and federated deployments, BeFaaS relies on unique function names, individual deployment artifacts for every function, and a single endpoint for every deployed function (\textbf{R3}):
With globally unique function names, the endpoints of the deployed functions are already known during the compilation phase.
The Deployment Compiler maps these endpoints to the canonical function names (defined in the application) and compiles them into the source code.
Moreover, the compiler also injects endpoints to external services such as database systems using environment variables which were set in the respective setup script or defined manually.
To enable asynchronous function calls, the Deployment Compiler creates and assigns a topic-based event pipeline on the respective provider for each asynchronous function.
To trigger this pipeline, requests are sent as events to a publisher function, which is deployed for every provider and forwards the request to the respective event topic (see \cref{fig:publisher}).
In total, this decouples the ability of a function to call another function or a platform service from its deployment location and enables BeFaaS to support arbitrarily complex deployments: it is indeed possible to run every function on a different FaaS platform -- as configured by the benchmarker.

Each FaaS platform offers a different interface for life-cycle and configuration management of functions.
As the smallest common interface, BeFaaS requires that each platform provides API-based access to (i)~deploying functions, (ii)~retrieving log entries from the standard logging interface, and (iii)~removing functions.
The Deployment Compiler wraps this functionality using an adapter mechanism and selects the appropriate instructions for the target platform specified in the deployment configuration.
Additional FaaS platforms that fulfill this minimal interface can easily be added by implementing a corresponding adapter (\textbf{R4}).

\subsection{Detailed Request Tracing}
To enable a detailed drill-down analysis of experiment results (\textbf{R5}), the Deployment Compiler injects and wraps code that collects detailed measurements during the benchmark run:
The compiler adds timestamping to determine start, end, and latency of calls to functions and external services.

Besides these timestamps, the compiler also injects code that generates context IDs and pair IDs to assign individual calls to their respective context later on.
Here, a context ID is generated once for each function chain (with the first function call) and propagated to every subsequent call to other functions.
To link the individual calls of a function chain, the compiler injects source code to create pair IDs of randomly generated keys that link caller and callee.
Thus, it is possible to trace every single request through the benchmark application and to generate call trees for every context and function chain during post-experiment analysis.

Finally, to independently and reliably detect cold starts, the Deployment Compiler also injects code that evaluates a local environment variable on the executor at the provider side.
If this variable is not present, the function runs on a new executor (cold start), the variable is created, filled with a randomly generated key, and the cold start is logged.

All data that enable fine-grained results (timestamps, context IDs, pair IDs, and executor keys) are recorded on the console using the standard logging interface of the respective FaaS vendor.
In initial experiments with AWS, GCP, and Azure, we verified that the cost of logging is at most in the microsecond range.

\subsection{Automated Experiment Orchestration\label{subsec:design-last}}
The BeFaaS framework requires only the application code, a deployment configuration, and a load profile to automatically perform the benchmark experiment (\textbf{R6}).
First, all business logic, dependencies, and BeFaaS instrumentation logic are bundled into a single deployment artifact by the Deployment Compiler.
Next, the Benchmark Manager orchestrates the experiment and provides a simple interface for starting the benchmark run, monitoring its process, and collecting fine-grained results for further analysis.

\section{Implementation}
\label{sec:impl}

Our open-source prototype implementation of BeFaaS\footnote{\url{https://github.com/Be-FaaS}} includes (i)~the BeFaaS library, (ii)~four deployment adapters, (iii)~the Deployment Compiler, (iv)~the Benchmark Manager, (v)~four realistic benchmark applications, and (vi)~several load profiles for the benchmark applications.

The BeFaaS library is written in JavaScript and handles calls to other functions depending on their canonical name, generates tracing IDs, and takes timestamps.
BeFaaS deployment adapters are implemented using Terraform\footnote{\url{https://www.terraform.io/}} commands.
Currently, BeFaaS thus supports three major cloud offerings (AWS Lambda, Google Cloud Functions, and Azure Functions) as well as the open-source system tinyFaaS~\cite{pfandzelter_tinyfaas_2020}, which supports the deployment of functions on private infrastructure, including edge or fog nodes.
The Deployment Compiler is a shell script that uses several tools to build the deployment adapters for the respective platforms, parses and injects information from the Deployment Configuration, and generates the deployment artifacts from the application code.
The Benchmark Manager uses Terraform to create the infrastructure based on these artifacts, collect the logs, and later remove provisioned resources.
The implemented benchmark applications are written in JavaScript and include calls to external services such as a Redis\footnote{\url{https://redis.io/}} instance.
The Load Generator uses Artillery\footnote{\url{https://artillery.io/}} to call the benchmark application.
It either executes a realistic default load profile that stresses all relevant aspects of the application or specific additional load profiles that emphasize stress situations, e.g., to provoke more cold starts.
New load profiles can easily be added by specifying new Artillery load descriptions (YAML\footnote{\url{https://yaml.org/}} configuration files).

\subsection{Benchmark 1: Web Shop (Microservices)}

\begin{figure}
    \centering
    \includegraphics[width=0.95\columnwidth]{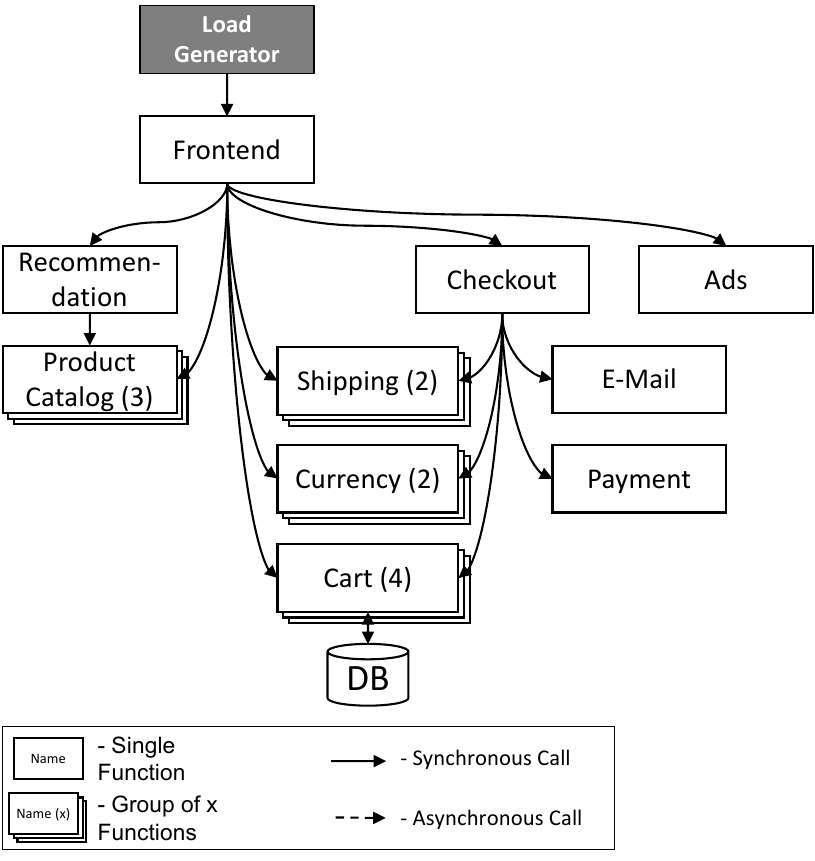}
    \caption{The e-commerce application implements a web shop in 17 functions. The \textsf{Frontend} serves as a single entry point and an external database is used to store state. We group some functions to increase legibility.}
    \label{fig:ecomm}
\end{figure}

Our e-commerce benchmark implements a web shop as a FaaS application derived from Google's microservice demo application.\footnote{\url{https://github.com/GoogleCloudPlatform/microservices-demo}}
Our corresponding benchmark implementation follows a typical request-response invocation style, comprises 17 functions, and uses a Redis instance as an external service to persist state (see \cref{fig:ecomm}).
Besides functions that provide recommendations and advertising, customers can log-in, set their preferred currency, view products, fill a virtual shopping cart, check out orders, and finally observe order shipping.
Each task is implemented in a separate function and all requests arrive at a single function, the frontend, which takes the customer calls and routes them to the respective backend functions.
There are blocking synchronous calls to other functions as well as asynchronous call blocks that idle until all called functions return.

The default load profile simulates four different customer workflows and constant traffic for 15~minutes.
Our e-commerce benchmark is particularly well suited for comparing request-response behavior and study request details of different cloud providers but can also be used to explore federated cloud deployments, e.g., for scenarios in which the application is running on multiple cloud platforms.

\subsection{Benchmark 2: Smart City (Hybrid Edge-Cloud)}

Although several IoT applications and use cases already exist in research (e.g.,~\cite{aral_learning_2020, brambilla_simulation_2014, ramprasad_smart_2018, mcchesney_defog_2019,grambow_public_2018}), none of them could directly be used or adapted as a FaaS application. 
Thus, we designed our smart city benchmark application around typical IoT patterns and implemented a use case based on a smart traffic control scenario inspired by the \emph{InTraSafEd5G} system~\cite{lujic_increasing_2021,paper_demaio_tarot_serverless}.

\begin{figure}
    \centering
    \includegraphics[width=0.95\columnwidth]{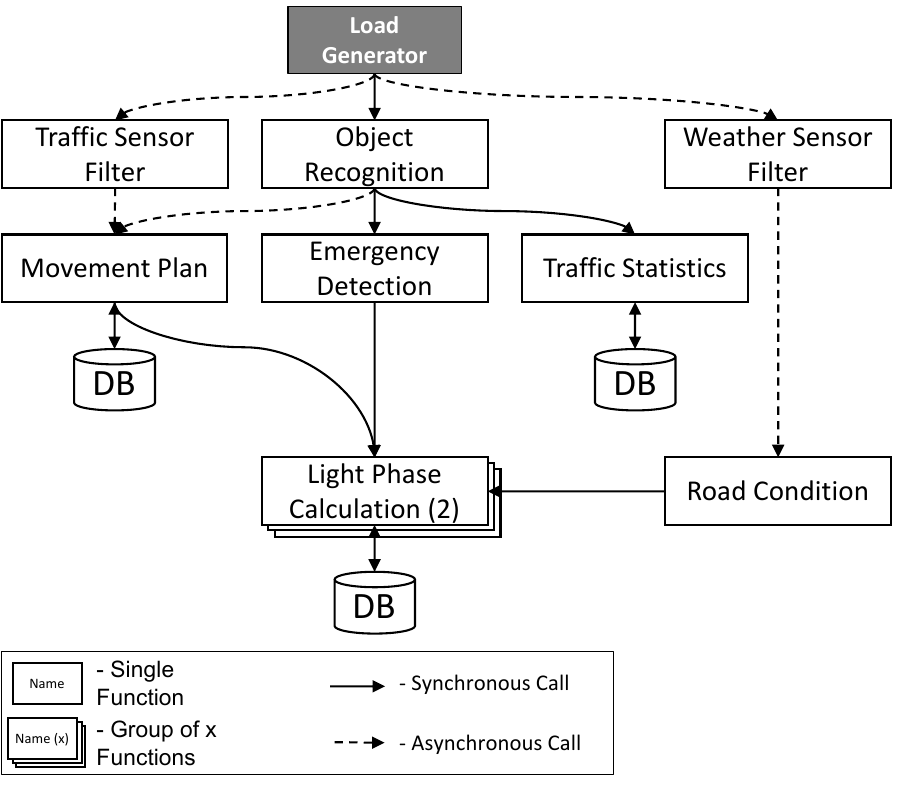}
    \caption{The IoT application implements a smart traffic light scenario in nine functions. The \textsf{Load Generator} emulates sensor data and sends them to three different entry points.}
    \label{fig:iot}
\end{figure}

The benchmark application uses a mix of synchronous and asynchronous function calls and implements an IoT use case with a smart traffic light which adapts its light phase based on traffic sensors, a camera, and weather inputs (see \cref{fig:iot}).
The functions initially filter incoming data streams and perform object recognition on camera footage to create a movement plan, detect ambulance/emergency cars, and maintain a traffic statistic.
The regular light phase is then determined based on this movement plan, road conditions, and the current light phase.
Emergency services can override the regular phase at any time by raising an emergency event that stops all other traffic.

The load profile for this application emulates sensor data and injects emergency events.
The traffic sensor sends an update every two seconds to the Traffic Sensor Filter, the Object Recognition processes one image every two seconds, and the weather is updated every twenty seconds.
Furthermore, the Load Generator also injects an emergency event every two minutes which lasts five seconds each.
This default load profile runs for 15 minutes.
As this use case will in practice typically have a very predictable and stable load profile, we did not implement alternative load profiles -- benchmark users can, however, easily add them if needed.

The smart city benchmark is particularly well suited for comparing different deployments across cloud, edge, and fog.

\subsection{Benchmark 3: Smart Factory (Event Trigger)}

\begin{figure}
    \centering
    \includegraphics[width=0.8\columnwidth]{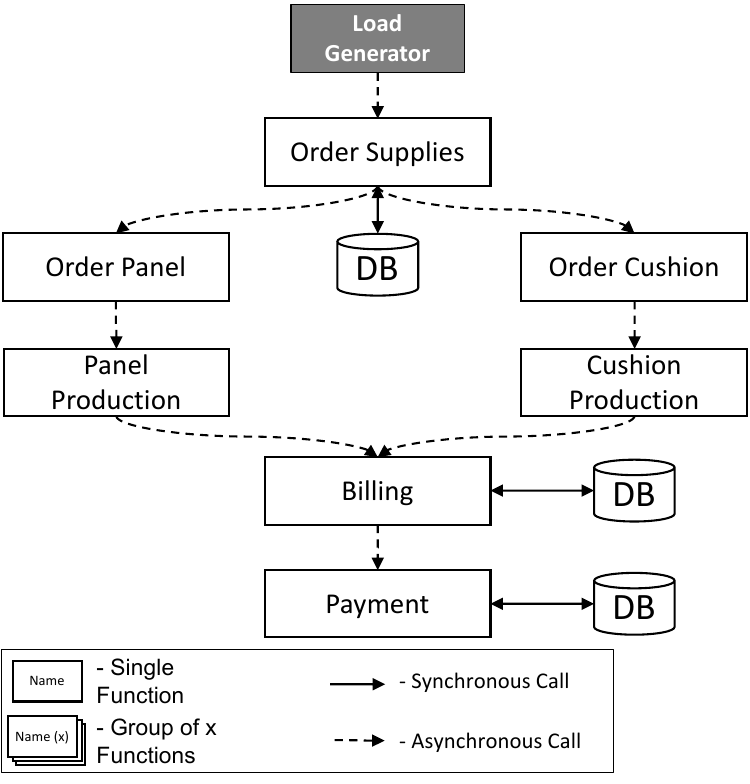}
    \caption{The Industry 4.0 application implements a smart factory in seven functions. An \textsf{Order Supplies} function serves as single entry point for different asynchronous event pipelines which can be distributed among several providers.}
    \label{fig:factory}
\end{figure}

Our smart factory benchmark application implements asynchronous event-based pipelines.
In our example use case, users order personalized couches consisting of panels and cushions (see \cref{fig:factory}).
First, an order function determines the number and individual sizes of panels and cushions which are then each ordered by sending an event to the respective order function.
Both order functions, in turn, transform their order into a production event which is sent to the production function which mimics the production of the panel or cushion.
After production, the functions emit an accounting event which is consumed by the billing function.
Once all panels and cushions are produced, and all accounting events are processed, the payment function finally issues an invoice.

The default load profile orders a new couch every five seconds for 15 minutes.
Each order, in turn, implies 8 to 18 order events, depending on the ordered couch model.
This smart factory benchmark application is particularly well suited to study event pipelines, but can also be used to analyze the interplay between different FaaS providers.
For example, when collaborating with suppliers (panels and cushions in our case), they may also be deployed on another provider, thus, requiring cross-cloud interoperability.

\subsection{Benchmark 4: Streaming Service (Cold Start Behavior)}

\begin{figure}
    \centering
    \includegraphics[width=0.95\columnwidth]{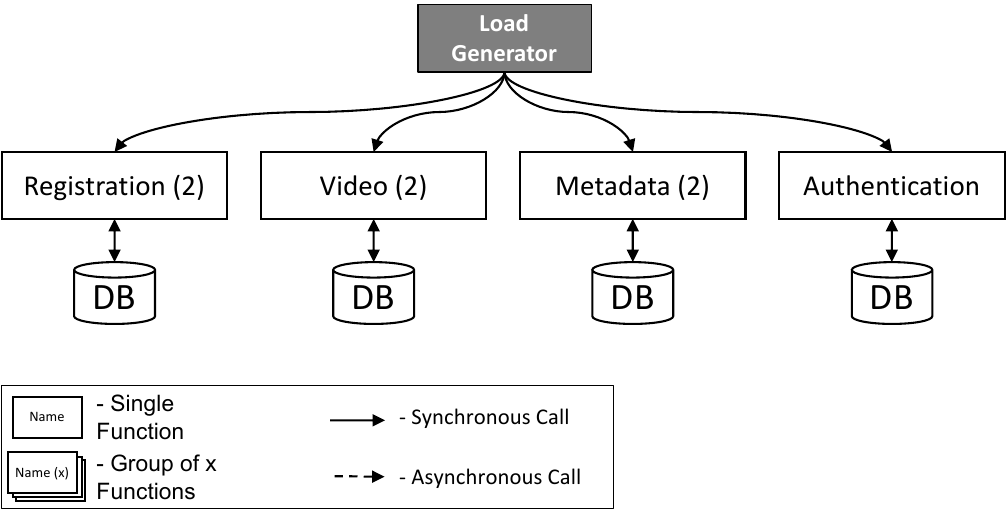}
    \caption{The streaming service comprises seven functions and a workload that triggers cold starts. After an Internet outage, there is a high load on functions dealing with authentication and metadata.}
    \label{fig:streaming}
\end{figure}

An often stated advantage of FaaS applications is their elastic scalability.
Thus, we include a streaming service benchmark application which triggers cold starts and can require automatic scaling capabilities.
The Load Generator here mimics video streaming devices which register users, request video files, update meta information such as viewing progress, and handle backend authentication (see \cref{fig:streaming}).
In case of a larger Internet outage, these devices are offline, but it is possible to continue watching already downloaded movies while the corresponding metadata is updated.
As soon as network connectivity is restored and the streaming devices are back online, all clients reconnect and concurrently invoke functions, which triggers cold starts.

The default load profile for this benchmark is split into four phases.
First, an initial set of users and streaming devices is registered.
Once all initial data has been read, the normal load phase starts for 5 minutes in which 500 request flows add new videos, request videos, and update metadata.
Third, the failure is simulated by pausing requests for 20 minutes.
Finally, the benchmark triggers cold starts by suddenly sending 1,500 request flows distributed over another 5 minutes.
Our streaming application benchmark is particularly well suited for comparing the cold start behavior and automatic scaling capabilities of different FaaS providers.

\section{Evaluation}
\label{sec:evaluation}
Our evaluation is split into two parts:
First, we present the results of four experiments in which we use BeFaaS to stress different FaaS platforms (\crefrange{subsec:experiment1}{subsec:experiment4}).
Second, in \cref{subsec:disc-reqs}, we discuss to which degree BeFaaS fulfills our requirements from \cref{sec:requirements}.

In all experiments, we deploy the Load Generator on a (vastly over-provisioned) virtual machine (2~vCPUs and 4~GB~RAM) and let it execute the default load profile of the respective benchmark application against the SUT deployed in either \textit{eu-west-1} for AWS, \textit{westeurope} for Azure, or \textit{europe-west1} for GCP.
As runtime for the benchmark applications, we run node.js 18 on all SUT options and use 256MB of memory per function (SKU Y1 on Azure).
Moreover, the Redis database system used by the SUTs also runs on an over-provisioned virtual machine (2~vCPUs and 4~GB~RAM; \textit{ta3.medium} at AWS, \textit{Standard\_B2S} in Azure, and \textit{e2-medium} at GCP) at the respective provider site.
This ensures that the database instance and Load Generator will not be a bottleneck during the experiments~\cite{book_cloud_service_benchmarking}.
All experiment results reported here are from the period June to July 2023.
We explicitly decided not to compare to the results from our original paper~\cite{grambow_befaas_2021} as we made a number of smaller changes across the BeFaaS codebase and also used the opportunity to update all libraries and platform SDKs used.
As a result, we reran all experiments from scratch since we could not rule out effects from our benchmarking tool.

\subsection{Experiment 1: Using the web shop application benchmark to compare major cloud FaaS providers\label{subsec:experiment1}}
\begin{figure}
    \centering
    \includegraphics[width=0.6\columnwidth]{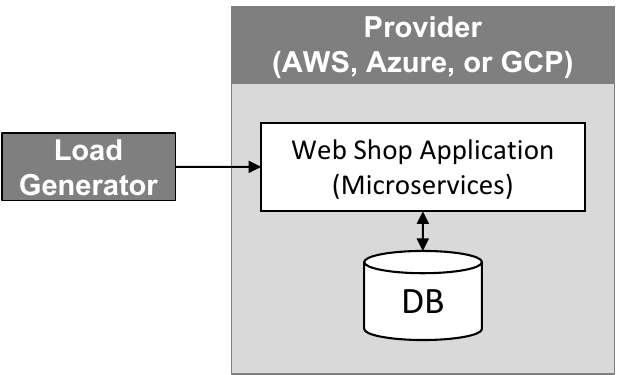}
    \caption{As part of the FaaS application, the database instance is deployed in the same region and on the same provider as the rest of the web shop.}
    \label{fig:eval1}
\end{figure}

In our first experiment, we deploy BeFaaS in single cloud provider setups in which all functions of the web shop application are deployed on a single provider (namely AWS, Azure, and GCP) and use the default load profile to compare them (see \cref{fig:eval1}).
During each experiment, the Load Generator executes $18,000$ workflows, which each consist of 1 to 9 requests, over a time span of 15 minutes.

\begin{figure}
    \centering
    \includegraphics[width=\columnwidth]{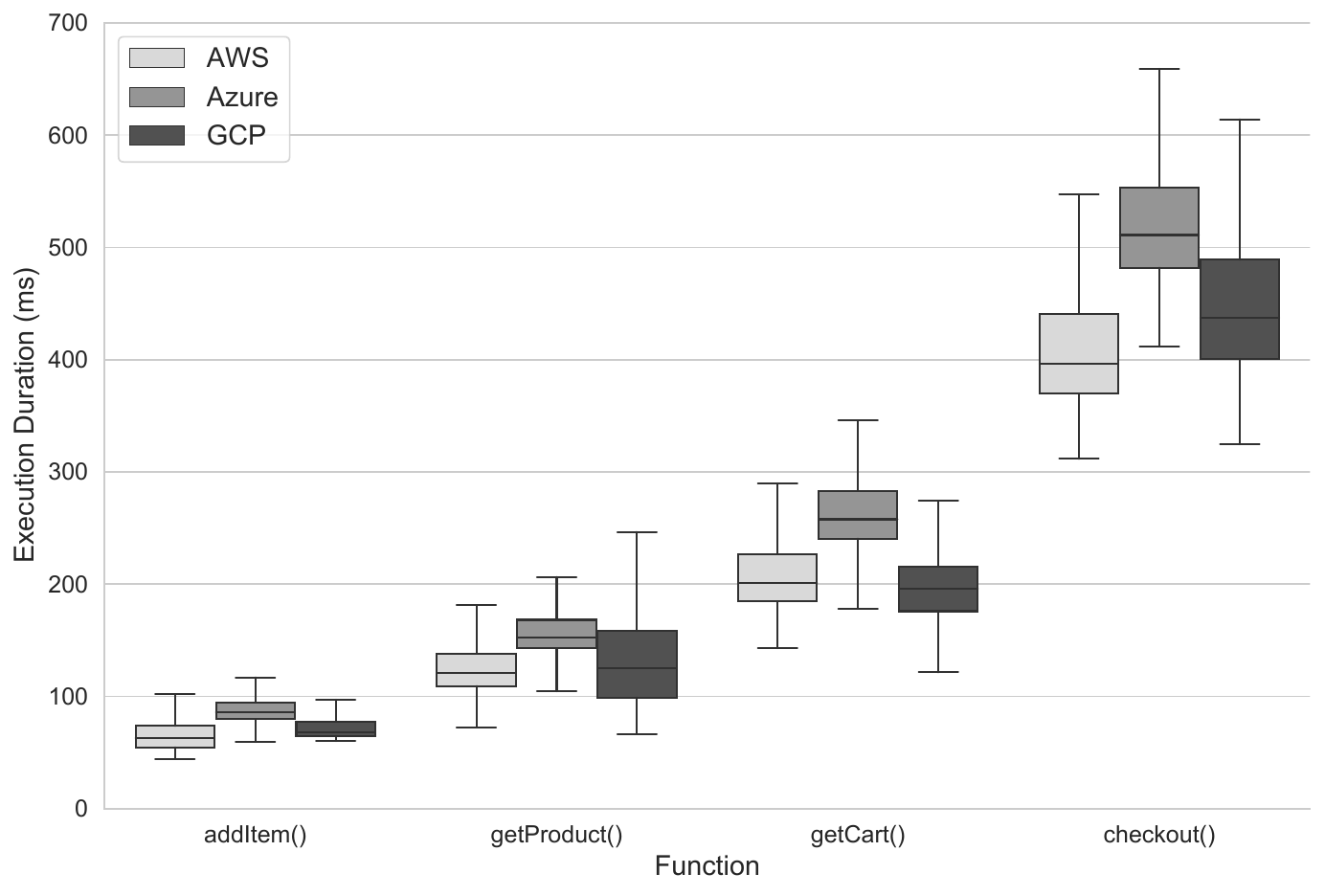}
    \caption{A detailed analysis of four functions called from the frontend shows that AWS provides the best performance and that the execution duration has the highest variance on GCP.}
    \label{fig:eval1result1}
\end{figure}

\Cref{fig:eval1result1} shows the execution duration of four selected functions with varying degree of complexity which are called from the frontend function (visualized as box plots; boxes represent quartiles, whiskers show the minimum and maximum values without outliers beyond 1.5 times the interquartile range).
For the four functions examined in more detail, the overall picture is similar for all three providers:
As expected, simpler functions that only read or write a single value have a lower execution duration than more complex ones such as the \textit{getCart()} or \textit{checkout()} function.
In our experiment, Azure provided the slowest environment for this single run while GCP showed a higher variance for the \textit{getProduct()} and \textit{checkout()} function.

In a further fine-grained analysis, we investigate the distribution of computing, network transmission, and database query latency for a function sequence putting an item into the shopping cart.
This includes synchronous and blocking calls to two functions and several database operations.

For this evaluation, we consider the (i)~computation part as function execution duration without the duration of outgoing network calls, (ii)~network latency as the duration of outgoing calls to other function without the execution duration of the called function itself, and (iii)~query latency as the duration of calls to the external database.
The detailed timestamp mechanisms of BeFaaS allow us to easily separate these times.

\begin{figure}
    \centering
    \includegraphics[width=\columnwidth]{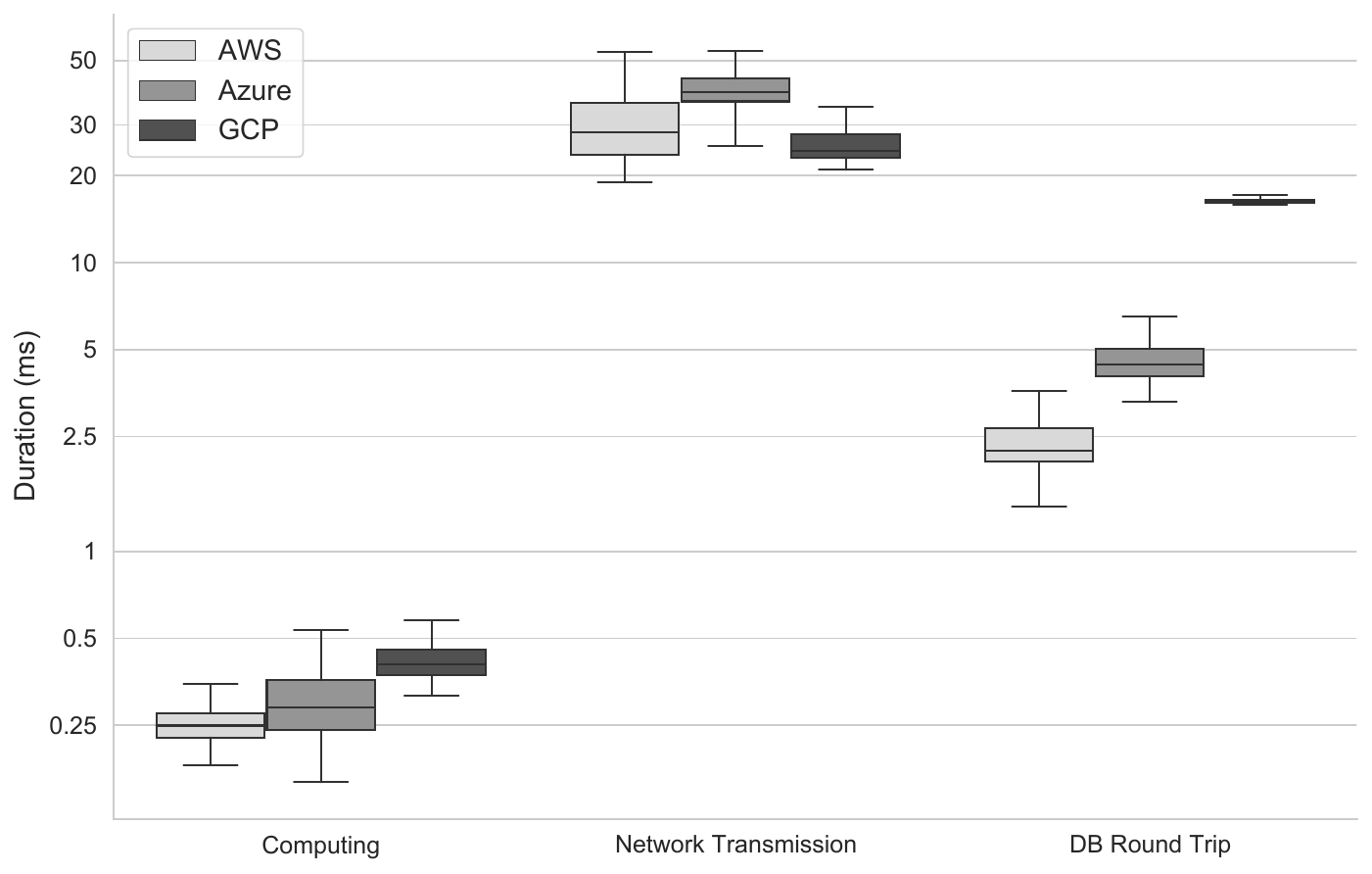}
    \caption{A drill-down analysis of a function sequence reveals that the network transmission time is the most relevant driver of execution time on all providers.}
    \label{fig:eval1result2}
\end{figure}

The results of this analysis are shown in \cref{fig:eval1result2}.
In this specific but typical interaction in which FaaS functions are the glue code to interact with external services, it is noticeable that for all providers time is mostly spent on network transmission followed by the database round-trip time while the actual computing time is below 1ms for all providers.
Furthermore, database access takes longer for GCP than for the other providers.

\subsection{Experiment 2: Evaluating hybrid edge-cloud setups using the smart city application\label{subsec:experiment2}}
\begin{figure}
    \centering
    \includegraphics[width=0.95\columnwidth]{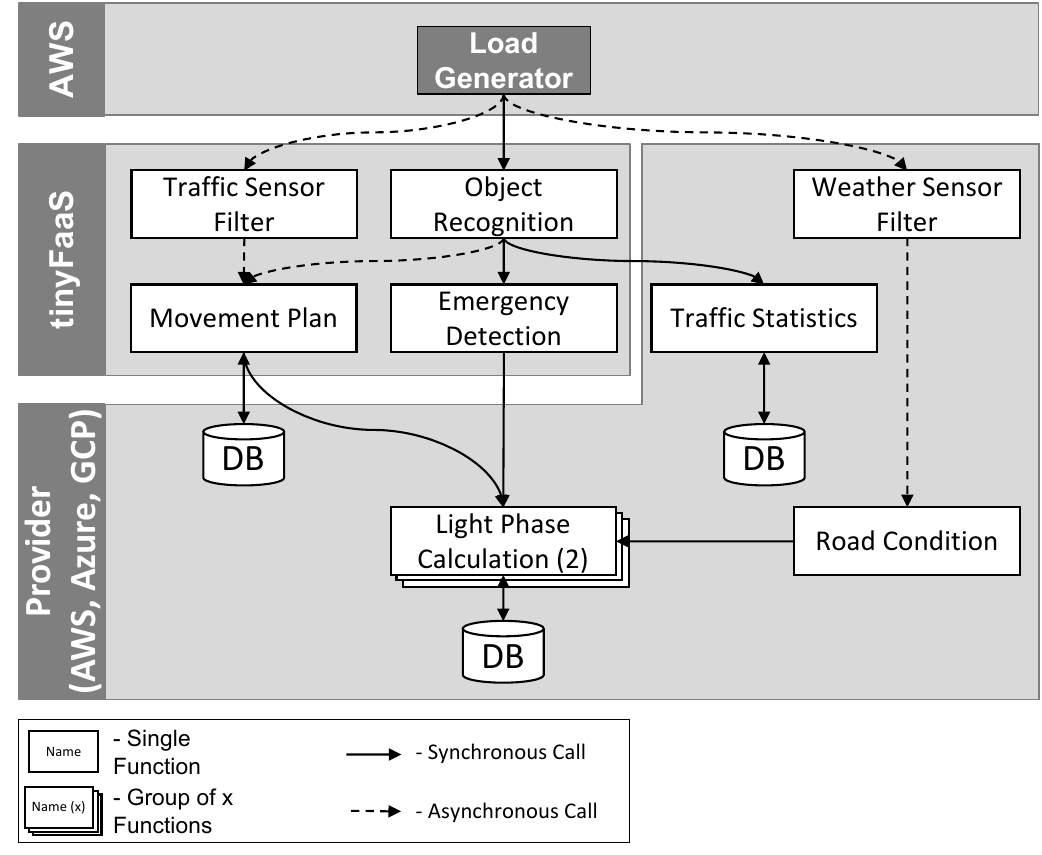}
    \caption{Functions related to the traffic light are deployed on a Raspberry Pi at the edge location while others run on public cloud providers.}
    \label{fig:eval2}
\end{figure}

In this experiment, we compare an edge-focused and a hybrid edge-cloud setup.
For the mixed setup, we split the smart city application into a cloud part, which is deployed on either AWS, Azure, or GCP, and an edge part, which is deployed on a local Raspberry Pi in Berlin, Germany running the tinyFaaS platform (see \cref{fig:eval2}).
For the edge-focused setup, we deploy all functions of the smart factory application on tinyFaaS and only use a cloud-located database at the respective provider.
During the experiment, the Load Generator simulates the smart city scenario for 15 minutes by triggering the traffic sensor and object recognition every two seconds and the weather sensor every 10 seconds for both evaluated setups.

\begin{figure}
    \centering
    \includegraphics[width=\columnwidth]{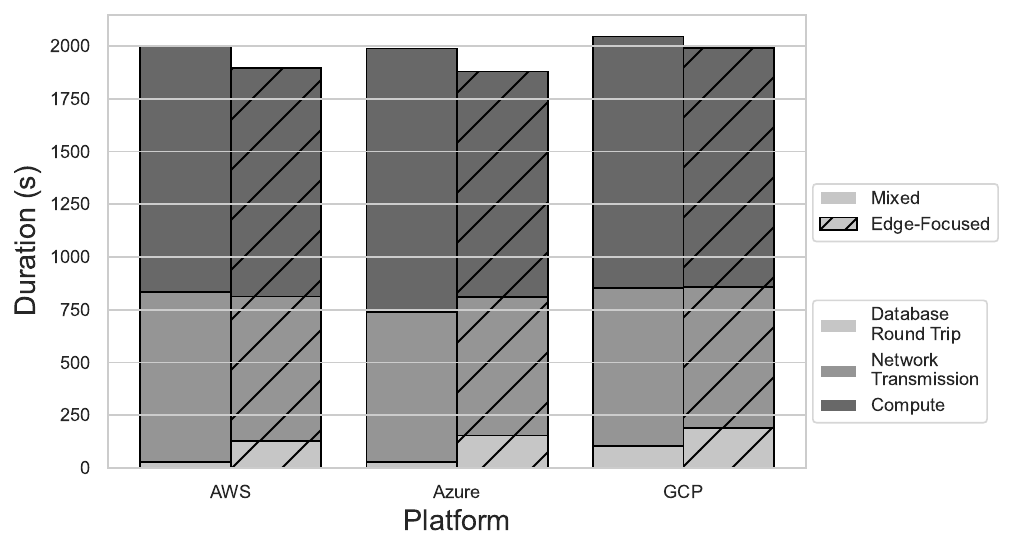}
    \caption{The cloud database increases the database duration for the edge setup for all providers. In total, however, both compute and network duration are reduced, and the total execution duration is lower for all edge setups.}
    \label{fig:eval2_results}
\end{figure}

Similar to our first experiment, we analyze the computing, network transmission, and database round trip times for both setups and all providers (see \cref{fig:eval2_results}).
For all providers, the default workload finishes sightly faster in the edge-focused setup.
Here, the network transmission times are reduced as all functions run on the same edge device, the compute duration is shorter compared to the mixed cloud-edge setup, but the database round trip time increases as every database access triggers a cloud request.
This last addition, however, does not outweigh the other two improvements for the given default load.

\subsection{Experiment 3: Analyzing the event pipeline interplay within and across FaaS providers.\label{subsec:experiment3}}
\begin{figure}
    \centering
    \includegraphics[width=0.7\columnwidth]{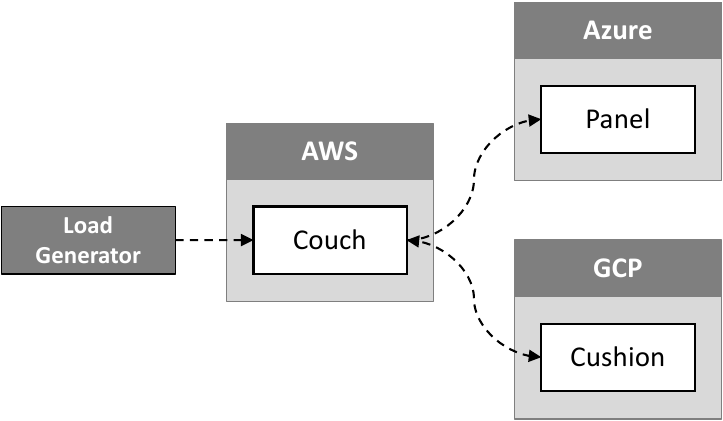}
    \caption{Each provider hosts a part of the smart factory application.}
    \label{fig:eval3}
\end{figure}

This experiment intends to investigate both how event pipelines perform within a provider, but also how well the interplay of cross-provider pipelines works.
Thus, we split the event-based smart factory application into three parts: 1.~\textsf{Couch} (running supplies, billing, and payment), 2.~\textsf{Panel} (running panel order and production), and 3.~\textsf{Cushion} (running cushion order and production).
Each part is deployed on AWS, Azure, or GCP (see \cref{fig:eval3}).
The load for this experiment consists of 180 couch orders over a time span of 15min, which triggers thousands of function invocations.

\begin{figure}
    \centering
    \includegraphics[width=\columnwidth]{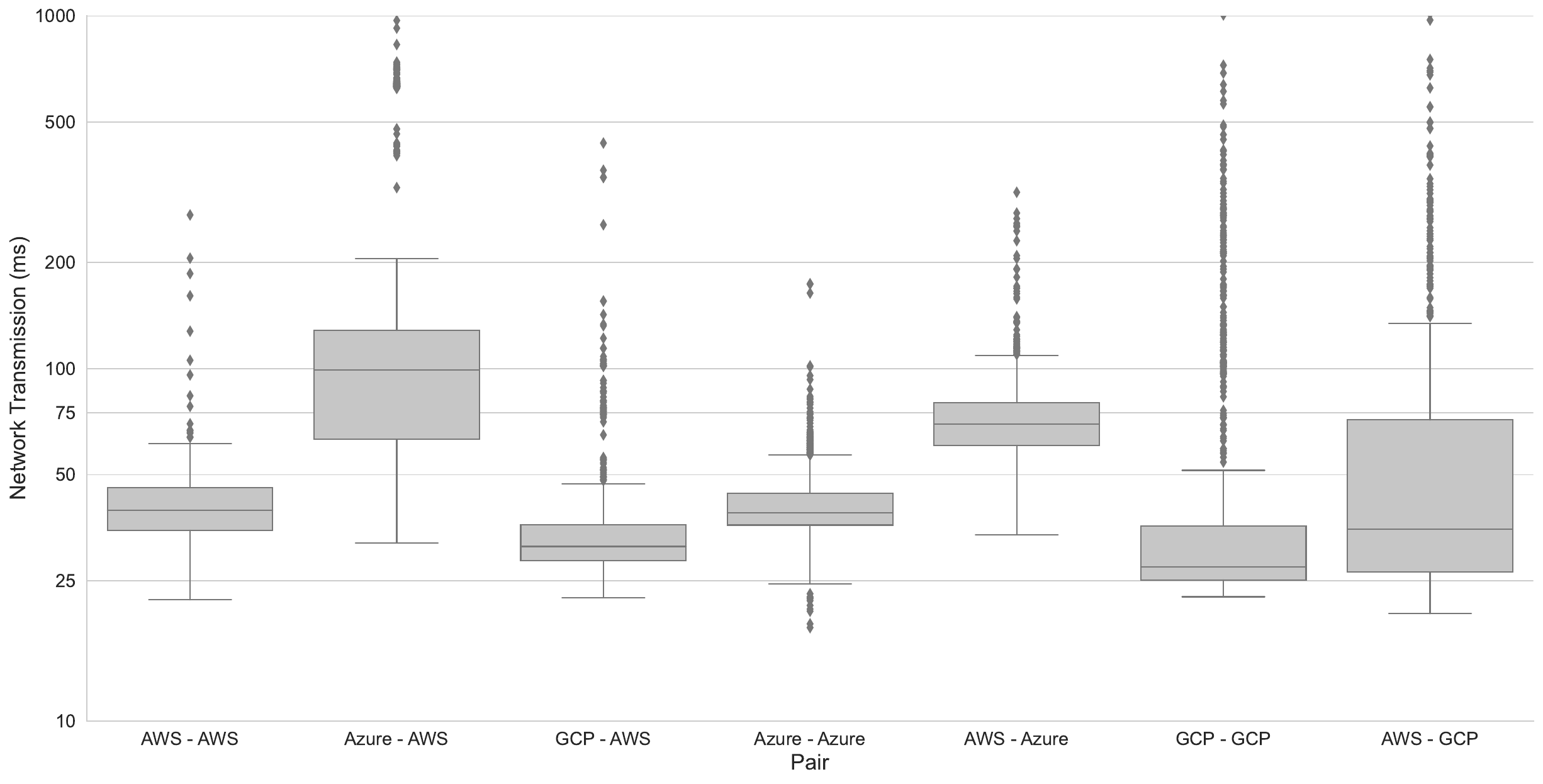}
    \caption{The network latency to publish an event usually ranges between 25ms and 100ms.}
    \label{fig:eval3_results_1}
\end{figure}

First, we analyze the time it takes to publish an event at the respective provider endpoint (see \cref{fig:eval3_results_1}).
Again, we measure the outgoing call from the calling function and subtract the execution duration from the execution time of the called publisher function running on the destination provider.
Depending on the origin and destination provider this usually takes between 25ms and 100ms, except for the Azure-AWS pair, which may take up to 200ms.
Moreover, there are outlier values of more than one second in five pairs.

\begin{figure}
    \centering
    \includegraphics[width=\columnwidth]{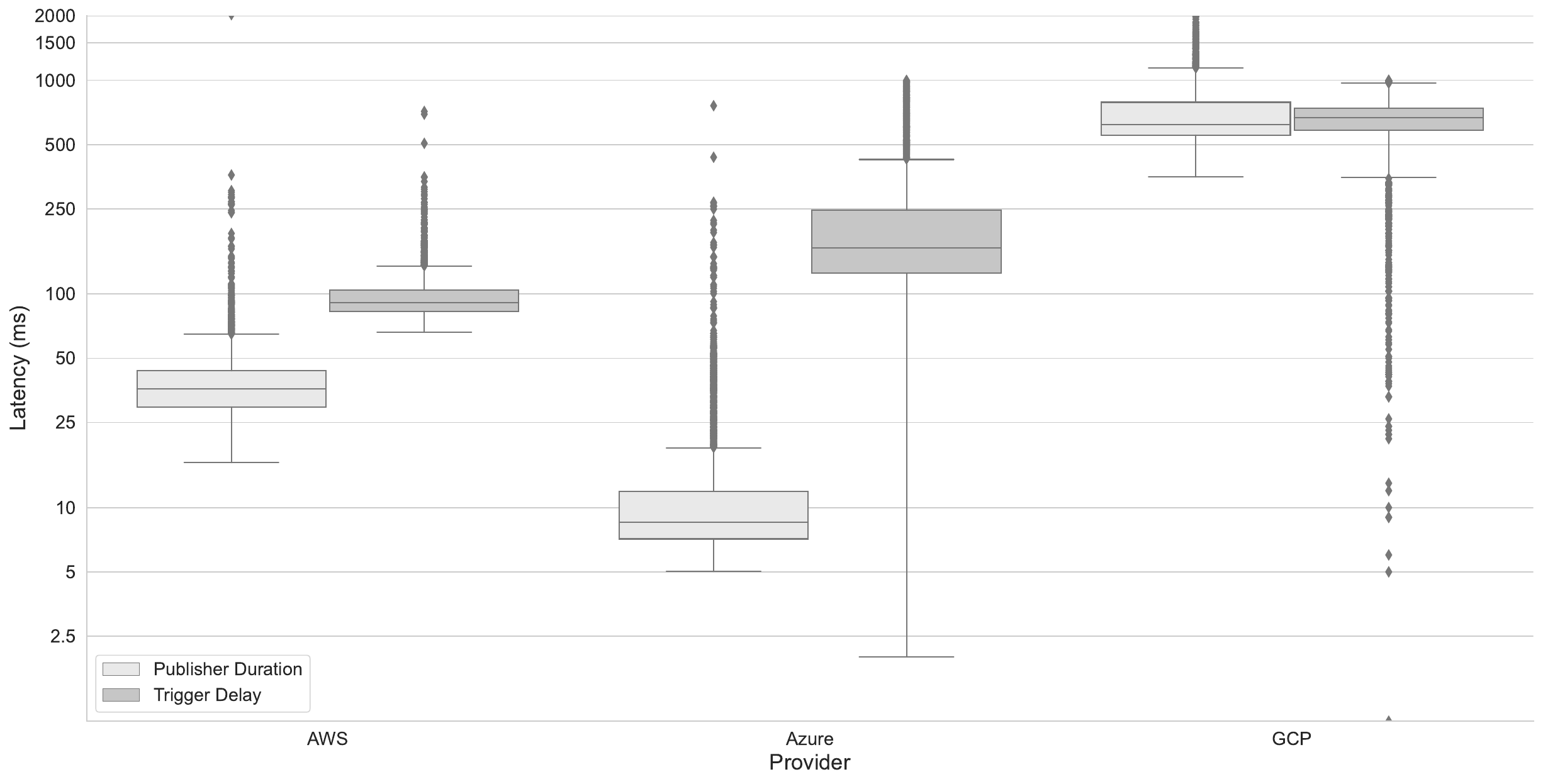}
    \caption{The execution time of the publisher function ranges from about 10ms on Azure to about 800ms on GCP. The trigger delay between publisher start and function start varies from 100ms on AWS to about 800ms on GCP.}
    \label{fig:eval3_results_2}
\end{figure}

Second, we also investigate the total execution time of the publisher function running on the destination provider and the trigger delay between the start of the publisher function and the start of the triggered function (see \cref{fig:eval3_results_2}).
Here, Azure and AWS show fast publishing functions which usually finish within 100ms while for GCP it usually takes between 500ms and 1000ms to run the publisher function.
For the trigger delay, AWS triggers the respective function fastest with about 100ms, followed by Azure with 75\% of values below 250ms, and GCP with 75\% of values above 500ms.
Furthermore, it is noticeable that outliers in the other direction are possible, i.e., an event sometimes triggers the function execution immediately within 10ms for Azure and GCP.

\subsection{Experiment 4: Studying the cold start behavior of different providers.\label{subsec:experiment4}}
\begin{figure}
    \centering
    \includegraphics[width=0.7\columnwidth]{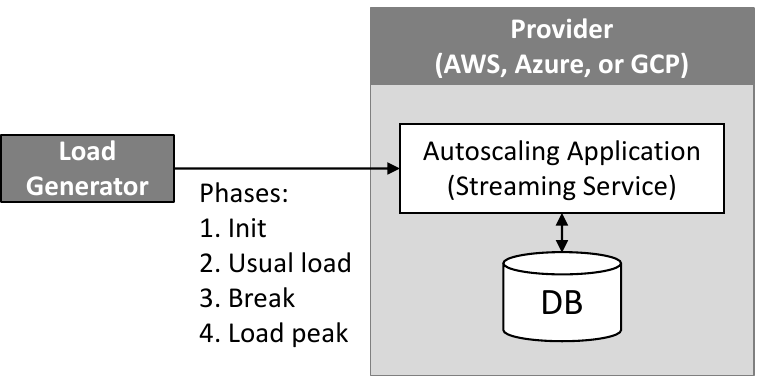}
    \caption{Similar to experiment 1, we deploy BeFaaS in single cloud provider setups in which all functions of the steaming service application are deployed on a single provider (namely AWS, Azure, and GCP).}
    \label{fig:eval4}
\end{figure}

In our last experiment, we study the cold start behavior of all cloud providers and deploy the streaming service application once on each provider (see \cref{fig:eval4}).
Running the default workload for this use case, we execute 500 requests within the first 5min load phase, which is then followed by a 20min break, and finally execute 1,500 requests for another 5 minutes.

\begin{figure*}[t]
    \centering
    \includegraphics[width=\textwidth]{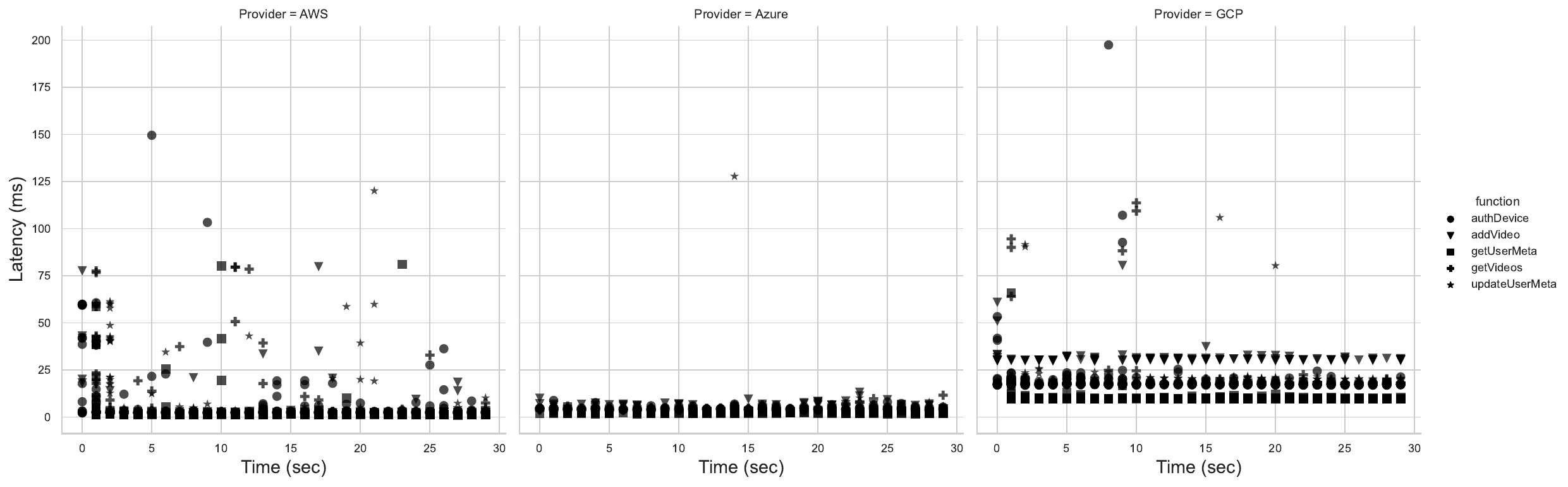}
    \caption{In the first 30 seconds of the load peak phase, AWS and GCP show larger latencies due to cold starts. Azure is presumably not affected by cold starts for this workload.}
    \label{fig:eval4_results_1}
\end{figure*}

\Cref{fig:eval4_results_1} shows the function execution time of all calls of the streaming application in the first 30 seconds of the \textit{Load Peak} phase.
In this phase, AWS and GCP both log cold starts and show larger latency values in the first seconds of the experiment.
While for AWS the values do not stabilize for the first 30 seconds, the GCP execution times are stable after 10 seconds.
Azure does not report any cold starts and shows stable execution durations.
However, we possibly also missed the respective log entry due to the log framework limitations:
The Azure logging framework only writes 250 log lines per second at maximum, further lines are discarded.
Thus, only 9,143 out of about 15,000 values across the whole experiment time are available for this analysis.

\subsection{Discussion of Requirements\label{subsec:disc-reqs}}
In \cref{sec:requirements}, we had identified six requirements for application-centric FaaS benchmarking frameworks.
We now discuss to which degree BeFaaS fulfills these requirements.

BeFaaS already comes with four standard benchmark applications that cover many representative FaaS application scenarios, namely standard web applications, a hybrid edge-cloud scenario, an event-based smart factory application, and a microservice-based streaming service.
Moreover, BeFaaS can be easily extended by implementing more FaaS application scenarios using the BeFaaS library and further workload profiles.
We, hence, believe that BeFaaS fulfills the requirements \textbf{R1} (\textit{Realistic Benchmark Application}) and \textbf{R2} (\textit{Extensibility for New Workloads}).

In BeFaaS, benchmark users can define arbitrarily complex deployment mappings of functions to target FaaS platforms including federated multi-cloud setups or mixed cloud/edge/fog deployments.
In fact, each function could run on a different platform. To achieve this, BeFaaS transforms the benchmark application into deployment artifacts fitted to the target platform.
Adding another target platform is also straightforward and only requires the benchmark user to implement an adapter component for the respective FaaS platform or to copy and adapt an existing adapter component.
Based on this, we argue that BeFaaS fulfills the requirements \textbf{R3} (\textit{Support for Modern Deployments}) and \textbf{R4} (\textit{Extensibility for New Platforms}).

At runtime, BeFaaS collects fine-grained measurements and traces individual requests similar to what Dapper~\cite{sigelman2010dapper} does for microservice applications.
This offers the necessary information basis for drill-down analysis.
Beyond this, BeFaaS also offers visualization capabilities for select standard measurements to further support analysis needs.
Overall, we hence conclude that BeFaaS addresses requirement \textbf{R5} (\textit{Support for Drill-down Analysis}).

Finally, we believe that BeFaaS is easy to use due to its experiment automation features and requires only very few configuration files (requirement \textbf{R6} -- \textit{Minimum Required Configuration Overhead}).
Nevertheless, this is a highly subjective matter that depends on the respective individual.
\section{Discussion}
\label{sec:discussion}
BeFaaS is a powerful application-centric FaaS benchmarking framework.
There are, however, also some points to consider and limits when using BeFaaS.

\paragraph*{Tracing Overhead}
BeFaaS supports a detailed tracing of requests by injecting a small token in each call.
On the one hand, this supports the clear mapping of different calls to function chains, yet on the other hand, it also causes an additional network overhead.
This token, however, is of constant size (depending on the length of the respective function name), so the overhead can be easily determined and considered in results analysis.
Furthermore, this will only matter if the goal of the benchmark is to find the optimal deployment for an existing application which is then instrumented to be used as a BeFaaS benchmark.
For any of our standard benchmarks, it simply increases the benchmark workload stress slightly.

\paragraph*{Measuring External Services}
Currently, BeFaaS handles external services and components as a black-box and only measures end-to-end latency of such service calls.
In future work, however, we plan to implement a small BeFaaS sidecar proxy that can be deployed on external service instances to forward calls to the respective service and to inject the BeFaaS tracing token there as well.

\paragraph*{Fairness with External Dependencies}
The included benchmark uses an external database system to persist state but further benchmarks and use cases may also require external services such as pub/sub message brokers or web APIs.
Although the modular design of BeFaaS supports this, there are also some pitfalls in terms of fairness and comparability:
In our experiments, we deployed the database instance with the same provider and in the same region as our functions to minimize latency between functions and database.
In this setup, a function calling the external service and awaiting the response will not idle for a long time and the execution environment at the provider side will soon be available again for the next request.
On the other hand, a function calling an external service in another region with larger latency will block the environment and (may) cause a cold start for the next incoming request.
Thus, when using external services, these should be located and deployed with similar latency for all alternatives.
Moreover, as cloud environments at least appear to be infinitely scalable, it has to be assured that the external service does not become a performance bottleneck during the experiment.
Otherwise, the benchmark would benchmark the compute resources of the external service instead of the performance of the FaaS platforms.

\paragraph*{Provider-specific Features}
Competing FaaS vendors are constantly developing new and exclusive features that simplify development and deployment for customers.
These features, however, can also affect the portability of the BeFaaS framework if a (future) benchmark uses exclusive features that are not available at all vendors.
Thus, we strongly recommend not to use exclusive features of individual providers when developing new BeFaaS benchmarks.
BeFaaS can, however, help to determine the impact of new features within a provider or across multiple providers by adjusting and configuring the respective deployment adapter.

\paragraph*{Time Synchronization}
The drill-down analysis features of BeFaaS require approximately synchronized clocks.
Although this will usually be provided by the provider with sufficient accuracy, a user should assert this before running experiments as this will affect the reliability of tracing insights.
Nevertheless, such detailed insights may often not be needed and the tracing of BeFaaS also offers a mechanism to partially mitigate this:
If the call follows a request-response pattern, BeFaaS measures the total round trip time at the calling function and knows the computing duration at the called function.
Thus, it is possible to approximate the network transmission latency under the assumptions that both directions took comparably long.
This is even possible for event-based calls that do not return a message to the sender, as calling functions submit the trigger events to the respective publisher function which runs on the same provider as the called function.
In our experience, though, this is neither a problem in the cloud nor for self-hosted FaaS platforms, where the user has direct control over clock synchronization.

\balance
\section{Conclusion}
\label{sec:conclusion}
FaaS platforms are a popular cloud compute paradigm and have also been proposed for edge environments.
For comparing and choosing different FaaS platforms in terms of performance, developers usually rely on benchmarking.
Existing FaaS benchmarks, however, tend to fall into the microbenchmark category -- an application-centric FaaS benchmarking framework is still missing.

In this paper, we presented BeFaaS, an extensible framework for executing application-centric benchmarks against FaaS platforms which comes with four realistic FaaS benchmark applications.
BeFaaS is the first benchmarking framework with out-of-the-box support for federated cloud setups which allows us to also evaluate complex configurations in which an application is distributed over multiple FaaS platforms running on a mixture of cloud, edge, and fog nodes.
Beyond this, BeFaaS is focused on ease-of-use through automation and collects fine-grained measurements which can be used for a detailed post-experiment drill-down analysis, e.g., to identify cold starts or other request-level effects; it can easily be extended with additional benchmarks or adapters for further FaaS platforms.

With BeFaaS, we provide developers with the necessary tool to explore, compare, and analyze FaaS platforms for their suitability for application scenarios.
We also offer researchers the ability to study the performance effects of different FaaS deployment options across cloud, edge, and fog through experiments.
Finally, FaaS platform developers can use BeFaaS in their CI/CD pipeline to compare their own platform to previous versions of it as well as to their competitors.

\section*{Acknowledgments}

We thank Luk Burchard, Emily Dietrich, Carsten Schubert, Christoph Witzko, and Max Zhao who contributed to the implementation of the first BeFaaS prototype within the scope of a master's project.
Moreover, we thank Luk Burchard, Carsten Schubert, and Max Zhao for also contributing to our initial paper.

\balance
\bibliographystyle{unsrt}
\bibliography{bibliography}

\vfill
\end{document}